\documentclass{article}
\begin{document}

\title{Classical and quantum Big Brake cosmology for scalar field and tachyonic models}

%\classification{(98.80.Qc, 04.60.Ds, 98.80.Jk }
%\keywords      {cosmology, singularities,scalar fields, tachyons}

\author{A.Yu. Kamenshchik$^{1}$ and S. Manti$^{2}$} 
%{
%  address={Dipartimento di Fisica e Astronomia and INFN, Via Irnerio 46, 40126 Bologna,
%Italy\\
%L.D. Landau Institute for Theoretical Physics of the Russian
%Academy of Sciences, Kosygin str. 2, 119334 Moscow, Russia}
%}

%\author{S. Manti}$^{2}$
%{
%  address={Scuola Normale Superiore, Piazza dei Cavalieri 7, 56126 Pisa, Italy}
%}

%\author{<author3>}{
%  address={<common address for author2 and author3>}
%  ,altaddress={<author1 address>} % additional visiting address
%}
\date{}
\maketitle
$^{1}$
Dipartimento di Fisica e Astronomia and INFN, Via Irnerio 46, 40126 Bologna,
Italy\\
L.D. Landau Institute for Theoretical Physics of the Russian
Academy of Sciences, Kosygin str. 2, 119334 Moscow, Russia\\
$^{2}$
Scuola Normale Superiore, Piazza dei Cavalieri 7, 56126 Pisa, Italy\\
%\maketitle
\begin{abstract}
We  study a  relation between the cosmological singularities in classical and quantum theory, comparing the classical and quantum dynamics in some models possessing the Big Brake singularity - the model based on a scalar field and two models based on a tachyon-pseudo-tachyon field .
It is shown that the effect of quantum avoidance is absent for the soft singularities of the Big Brake type while 
it is present for the Big Bang and Big Crunch singularities. Thus, there is some kind of a classical - quantum 
correspondence, because  soft singularities are traversable in classical cosmology, while the strong Big Bang and Big Crunch 
singularities are not traversable. 
\end{abstract}

%%%%%%%%%%%%%%%%%%%%%%%%%%%%%%%%%%%%%%%%%%%%
%% MAINMATTER
%%%%%%%%%%%%%%%%%%%%%%%%%%%%%%%%%%%%%%%%%%%%

\section{Introduction}
The cosmological singularities constitute one of the main problems of modern cosmology.
 ``Traditional'' or ``hard'' singularities such as the Big Bang and the Big Crunch singularities are associated with the zero volume of the universe 
(or of its scale factor), and with infinite values of the Hubble parameter, of the energy density and of the pressure.
The discovery of the cosmic acceleration stimulated the development of ``exotic'' cosmological models of dark energy; some of these models possess the so called soft or sudden singularities characterized by the finite value of the radius of the universe and of its 
Hubble parameter. Such singularities sometimes arise quite unexpectedly in some dark energy models. One of examples of such singularities is the Big 
Brake singularity arising in a specific tachyon model \cite{tach}.
Tachyons (Born-Infeld fields) is a natural candidate for a dark energy because their pressure is negative. 
The toy tachyon model \cite{tach}, proposed in 2004, has two particular features:
tachyon field transforms itself into a pseudo-tachyon field;
the evolution of the universe can encounter a new type of singularity - the Big Brake singularity. 

The Big Brake singularity is a particular type of the so called ``soft'' cosmological singularities - 
the radius of the universe is finite, the velocity of expansion is equal to zero, the deceleration is infinite.
Amusingl, the predictions of the model do not contradict observational data on supenovae of the type Ia \cite{tach1,tach2}
and moreovoer, the Big Brake singularity is indeed a very special  one - it is possible to cross it \cite{tach2}. 

Naturally some open questions arise: what can we say about other soft singularities - is it possible to cross them ?
An attempt to study this question was undertaken in \cite{tach3}.
Antoher question is as follows: What can tell us the 
Quantum cosmology on the Big Brake singularity and other soft singularities ?  This question was studied in \cite{we} 
and here we present briefly the main results of this work.

\section{Description of the tachyon model}
We consider a flat Friedmann universe
\begin{eqnarray*}
&&ds^2 = dt^2 - a^2(t)dl^2
\end{eqnarray*}
The tachyon Lagrange density is 
\begin{eqnarray*}
L = - V(T)\sqrt{1-\dot{T}^2}
\end{eqnarray*}
The energy density
\begin{eqnarray*}
\rho =  \frac{V(T)}{\sqrt{1-\dot{T}^2}}
\end{eqnarray*}
The pressure 
\begin{eqnarray*}
p = - V(T)\sqrt{1-\dot{T}^2}
\end{eqnarray*}
The Friedmann equation is as usual
\begin{eqnarray*}
H^2 \equiv \frac{\dot{a}^2}{a^2} = \rho
\end{eqnarray*}
The equation of motion for the tachyon field is 
\begin{eqnarray*}
&&\frac{\ddot{T}}{1-\dot{T}^{2}}+3H\dot{T}+\frac{V_{,T}}{V}=0.
\end{eqnarray*}
In our model \cite{tach}
\begin{eqnarray*}
&&V(T)=\frac{\Lambda }{\sin ^{2}\left[ \frac{3}{2}{\sqrt{\Lambda \,(1+k)}\ T}%
\right] }\\
&&\times\sqrt{1-(1+k)\cos ^{2}\left[ \frac{3}{2}{\sqrt{\Lambda \,(1+k)}\,T}%
\right] }\ ,  \label{VTfixed}
\end{eqnarray*}
where $k$ and $\Lambda > 0$ are the parameters of the model.
The case $k > 0$ is more interesting. In this case
some trajectories (cosmological evolutions) finish in the infinite de Sitter expansion.
In other trajectories the tachyon field transforms into the pseudotachyon field
with the Lagrange density, energy density and positive pressure given by 
\begin{eqnarray*}
&&L = W(T)\sqrt{\dot{T}^2-1},\\
&&\rho = \frac{W(T)}{\sqrt{\dot{T}^2-1}},\\ 
&&p = W(T)\sqrt{\dot{T}^2-1},\\
&&W(T) = \frac{\Lambda }{\sin ^{2}\left[ \frac{3}{2}{\sqrt{\Lambda \,(1+k)}\ T}%
\right] }\\
&&\times\sqrt{(1+k)\cos ^{2}\left[ \frac{3}{2}{\sqrt{\Lambda \,(1+k)}\,T}-1
\right] }.
\end{eqnarray*}
What happens with the Universe after the transformation of the tachyon into the pseudotachyon 
? It encounters the Big Brake cosmological singularity.

\section{The Big Brake cosmological singularity and other soft singularities}
The Big Brake singularity is characterized by the following formulae:
\begin{eqnarray*}
t \rightarrow t_{BB} < \infty
\end{eqnarray*}
\begin{eqnarray*}
a(t \rightarrow t_{BB}) \rightarrow a_{BB} < \infty 
\end{eqnarray*}
\begin{eqnarray*}
\dot{a}(t \rightarrow t_{BB}) \rightarrow 0
\end{eqnarray*}
\begin{eqnarray*}
\ddot{a}(t \rightarrow t_{BB}) \rightarrow -\infty
\end{eqnarray*}
\begin{eqnarray*}
R(t \rightarrow t_{BB}) \rightarrow +\infty
\end{eqnarray*}
\begin{eqnarray*}
T(t \rightarrow t_{BB}) \rightarrow T_{BB},\ |T_{BB}| < \infty
\end{eqnarray*}
\begin{eqnarray*}
|\dot{T}(t \rightarrow t_{BB})| \rightarrow \infty
\end{eqnarray*}
\begin{eqnarray*}
\rho(t \rightarrow t_{BB}) \rightarrow 0
\end{eqnarray*}
\begin{eqnarray*}
p(t \rightarrow t_{BB}) \rightarrow +\infty
\end{eqnarray*}
If $\dot{a}(t_{BB}) \neq 0$ it is a more general soft singularity.

\section{Crossing the Big Brake singularity and the future of the universe}
At the Big Brake singularity the equations for geodesics are regular, because 
the Christoffel symbols are regular (moreover, they are equal to zero).
Is it possible to cross the Big Brake ?
Let us study the regime of approaching the Big Brake.
Analyzing the equations of motion we find that approaching the Big Brake singularity  the tachyon field behaves as
\begin{eqnarray*}
T=T_{BB}+\left( \frac{4}{3W(T_{BB})}\right) ^{1/3}(t_{BB}-t)^{1/3}.
\label{tachBB}
\end{eqnarray*}
Its time derivative $s \equiv \dot{T}$
behaves as 
\begin{eqnarray*}
s=-\left( \frac{4}{81W(T_{BB})}\right) ^{1/3}(t_{BB}-t)^{-2/3},  \label{sBB}
\end{eqnarray*}
the cosmological radius is 
\begin{eqnarray*}
a=a_{BB}-\frac{3}{4}a_{BB}\left( \frac{9W^{2}(T_{BB})}{2}\right)
^{1/3}(t_{BB}-t)^{4/3},  \label{cosmradBB}
\end{eqnarray*}
its time derivative is
\begin{eqnarray*}
\dot{a}=a_{BB}\left( \frac{9W^{2}(T_{BB})}{2}\right) ^{1/3}(t_{BB}-t)^{1/3}
\label{cosmradderBB}
\end{eqnarray*}
and the Hubble variable is
\begin{eqnarray*}
H=\left( \frac{9W^{2}(T_{BB})}{2}\right) ^{1/3}(t_{BB}-t)^{1/3}.
\label{HubbleBB}
\end{eqnarray*}
All these expressions can be continued in the region where $t>t_{BB}$,which
amounts to crossing the Big Brake singularity. Only the expression for $s$
 is singular but this singularity is integrable and
not dangerous.
Once reaching the Big Brake, it is impossible for
the system to stay there because of the infinite
deceleration, which eventually leads to the decrease
of the scale factor. This is because after the Big
Brake crossing the
time derivative of the
cosmological radius  and Hubble
variable  change 
their signs. The expansion is then followed
by a contraction, culminating in the Big Crunch singularity.

\section{The classical and quantum dynamics in the scalar field model with a soft singularity}
One of the simplest cosmological models revealing the Big Brake singularity is the model based on the anti-Chaplygin gas with an equation of state 
\begin{eqnarray*}
p = \frac{A}{\rho},\ \ A > 0.
\end{eqnarray*}
Then
\begin{eqnarray*}
\rho(a)=\sqrt{\frac{B}{a^6}-A}
\end{eqnarray*}
At $a = a_* = \left(\frac{B}{A}\right)^{1/6}$ the universe encounters the Big Brake singularity.
The scalar field model reproducing the cosmological evolution of the model based on the 
anti-Chaplygin gas has the potential \cite{Kiefer}
\begin{eqnarray*}
V(\varphi) = \pm\frac{\sqrt{A}}{2}\left(\sinh 3\varphi -\frac{1}{\sinh 3\varphi}\right).
\end{eqnarray*}
We shall study the model with a more simple potential, which has basically the same qualitative behaviour
\begin{eqnarray*}
V = -\frac{V_0}{\varphi},\ \ V_0 > 0
\end{eqnarray*}
We shall study first the classical dynamics of this model.
Here are the main results of our analysis.
\begin{enumerate}
\item
The transitions between the positive and negative values of the scalar field are impossible.
\item
All the trajectories (cosmological evolutions) with positive values of the scalar field 
begin in the Big Bang singularity, then achieve a point of maximal expansion, then contract and 
end their evolution in the Big Crunch singularity.
\item   
All the trajectories with positive values of the scalar field pass through the point where the value of the scalar field 
is equal to zero. After that the value of the scalar field begin growing. The point $\varphi = 0$ corresponds to a crossing of the soft singularity.
\item 
If the moment when the universe achieves the point of the maximal expansion coincides with the moment of the  crossing of the soft singularity  then the singularity is the Big Brake. 
\end{enumerate}
For completeness we can add that
the evolutions with the negative values of the scalar field belong to two classes - first, an infinite expansion beginning from 
the Big Bang 
and second, the evolutions obtained by the time reversion of those of the first class, which are contracting and end 
in the Big Crunch singularity.

\section{Quantum dynamics - the Wheeler - DeWitt equation}
Applying the Hamiltonian formalism to the system Gravity + Scalar field, we obtain the super-Hamiltonian constraint
\begin{eqnarray*}
-\frac{p_a^2}{4a}+\frac{p_{\varphi}^2}{2a^3}+Va^3 = 0,
\label{constraint}
\end{eqnarray*}
and the implementation of the Dirac quantization procedure gives the Wheeler-DeWitt equation
\begin{eqnarray*}
\left(-\frac{\hat{p}_a^2}{4a}+\frac{\hat{p}_{\varphi}^2}{2a^3}+Va^3\right)\psi(a,\varphi) = 0,
\label{WDW}
\end{eqnarray*}
 \begin{eqnarray*}
\left(\frac{a^2}{4}\frac{\partial^2}{\partial a^2} - \frac12\frac{\partial^2}{\partial \varphi^2} -\frac{a^6V_0}{\varphi}\right)\psi(a,\varphi) = 0.
\label{WDW2}
\end{eqnarray*}
Requiring the normalizability of the wave function of the universe we come to 
the conclusion that this wavefunction should vanish at $\varphi \rightarrow 0$.
This value classically corresponds to a soft singularity.
Does it indicate the presence of a quantum avoidance of the singularity ? 
No.
The physical sense has the wave function depending on the physical degrees of freedom,
obtained after the gauge fixing choice, which simultaneously introduces the time parameter 
\cite{barvin}.
If we choose the Hubble parameter as a new time parameter $\tau \equiv -H$, then its conjugated is $a^3$. 
The  reduction of the initial set of variables to the smaller set of physical degrees of freedom implies the appearance of the Faddeev-Popov determinant which  is equal to the Poisson bracket of the gauge-fixing condition 
and the constraint.
This Faddeev-Popov determinant will be  proportional to the potential, which is singular 
 at $\varphi = 0$. 
 Thus,  the quantum probability to find the universe, crossing the soft singularity is different 
 from zero. 

The ``hard'' Big Bang and Big Crunch singularities $a = 0$ correspond to 
$\varphi = \infty$. 
To provide the normalizability of the wave function  one should have the integral on the values of the scalar field $\varphi$ convergent, when $|\varphi| \rightarrow \infty$. 
That means that, independently of details connected with the gauge choice, not only the wave function of the universe but also  the probability density of scalar field values  should decrease 
rather rapidly when the absolute value of the scalar field is increasing.\\ Thus, in this case, the effect of the quantum avoidance of the classical singularity is present. 

\section{The quantum cosmology of the tachyon model}
The Wheeler-DeWitt equation for the system Gravity + Pseudotachyon is
\begin{eqnarray*}
\left(\sqrt{\hat{p}_T^2-a^6W^2} - \frac{a^2\hat{p}_{a}^2}{4}\right)\psi(a,T) = 0.
\label{WDWTp}
\end{eqnarray*}
Let us first consider the case of the constant potential $W(T) = W_0$.
The Big Brake singularity corresponds to $|p_T| = a^3\sqrt{W_0}$.
It is convenient to work in the momenta representation $\psi(a,p_T)$.
Then requirement of the well-definiteness of 
\begin{eqnarray*}
\sqrt{\hat{p}_T^2-a^6W_0^2}\psi(a,p_T)
\end{eqnarray*}
becomes algebraic and implies  
\begin{eqnarray*}
\psi(a,p_T)|_{|p_T| =a^3\sqrt{W_0}} = 0.
\end{eqnarray*}
It does not mean that the probability of finding of the universe crossing the Big Brake is equal to zero because the corresponding Faddeev-Popov determinant contains a singular factor 
\begin{eqnarray*}
\sim \frac{1}{\sqrt{p_T^2-a^6W_0^2}}.
\end{eqnarray*}
One can see that in the case of the trigonometrical potential there is no need to require 
even the disappearance of $\psi(a,T)$ at the Big Brake. 
Thus, there is no a quantum avoidance effect of the Big Brake singularity in the tachyon models.
There is the effect of quantum avoidance for the Big Bang and Big Crunch singularities,
because these singularities correspond to 
\begin{eqnarray*}
|p_T| \rightarrow \infty
\end{eqnarray*}
and the corresponding probability density should tend to zero.

\section{Conclusions and discussion}
It wass shown that the effect of quantum avoidance is absent for the soft singularities of the Big Brake type while 
it is present for the Big Bang and Big Crunch singularities. 
Thus, there is some kind of a classical - quantum 
correspondence, because  soft singularities are traversable in classical cosmology, while the strong Big Bang and Big Crunch 
singularities are not traversable.

%\begin{theacknowledgments}
We are grateful to A.O. Barvinsky and C. Kiefer for fruitful discussions and to P.V. Moniz and M. Bouhmadi-Lopez for useful 
correspondence.
This work was partially supported by the RFBR grant 11-02-00643.  
%\end{theacknowledgments}

%\endinput
\end{document}